# A Comprehensive Approach to Ensuring Quality in Spreadsheet-Based Metadata

Martin J. O'Connor, Marcos Martínez-Romero, Mete Ugur Akdogan, Josef Hardi, Mark A. Musen

Stanford Center for Biomedical Informatics Research
Stanford, CA 94304, USA

**Abstract**

While scientists increasingly recognize the importance of metadata in describing their data, spreadsheets remain the preferred tool for supplying this information despite their limitations in ensuring compliance and quality. Various tools have been developed to address these limitations, but they suffer from their own shortcomings, such as steep learning curves and limited customization. In this paper, we describe an end-to-end approach that supports spreadsheet-based entry of metadata while providing rigorous compliance and quality control. Our approach employs several key strategies, including customizable templates for defining metadata, integral support for the use of controlled terminologies when defining these templates, and an interactive Web-based tool that allows users to rapidly identify and fix errors in the spreadsheet-based metadata they supply. We demonstrate how this approach is being deployed in a biomedical consortium to define and collect metadata about scientific experiments.

**Introduction**

Metadata, which provide descriptive information about data, are an essential component of the scientific endeavor. Accurate and comprehensive metadata are crucial to meet the goals of data FAIRness [1] and to help ensure that data can be understood, analyzed, and reused by other researchers. Many communities have created standardized reporting guidelines that specify the domain-specific metadata required to meet these goals. To further ensure high levels of standardization, use of controlled terminologies and ontologies is common in these guidelines.

While such reporting guidelines provide a core foundation for the creation of high-quality metadata, there are many practical challenges when researchers attempt to author metadata that conform to the guidelines. A key challenge is providing capabilities for authoring rich, standards-adherent metadata that fit with existing scientific workflows and tool choices. In particular, solutions that interoperate seamlessly with spreadsheets—overwhelmingly the most popular data-entry instrument for many researchers—can help address these challenges.

Scientists use spreadsheets to acquire data and metadata, to exchange these data and metadata with collaborators, to analyze them in tools such as Excel, and to access data and metadata in analysis pipelines that work directly with tabular entries. While spreadsheets are powerful tools for data management, they are not very good at enforcing compliance with community standards. Compliance errors can include missing required fields, typos, formatting errors, or values that do not conform to pre-specified value lists in the metadata specification. While some validation within spreadsheet tools is possible, is it quite limited. Tools such as Excel, for example, do allow users to set rules for what type of data is allowed in a cell or in a range of cells. Excel dropdowns can also be used to suggest a set of pre-defined values for some cells, which can encourage users to supply correct values from pre-specified lists.

While these constraints can encourage compliance to metadata specifications, they cannot ensure it. Users are still free to blithely ignore the constraints and to supply erroneous metadata. As a result, metadata ingestion processes must anticipate and handle a large variety of possible errors in user-submitted metadata spreadsheets. Fixing many of these errors may require manual intervention from curators, which may require contacting submitters to help rectify those errors—assuming that the curators are able to detect the errors in the first place. The end result can be a poor user experience for metadata submitters and expensive-to-maintain ingestion pipelines, all of which can lead to poor quality metadata.

There is a pressing need for tools that can facilitate the creation of high-quality, standards-adherent metadata and that can still allow researchers to use their familiar spreadsheets. In this paper, we outline such a system. Building on an existing metadata management platform called CEDAR (https://metadatacenter.org) [2], we outline how we have developed an end-to-end solution for encoding metadata specifications and then ensuring strong compliance to those specifications. Our approach employs customizable templates for defining metadata, integral support for the use of controlled terminologies, and an interactive Web-based tool for metadata validation and repair. We show how the resulting system provides metadata specification, acquisition, validation, and repair capabilities that can help ensure high-quality metadata. We demonstrate how our approach is being deployed in a biomedical consortium to define and collect metadata about scientific experiments, and how it provides a more effective and efficient solution for ensuring high-quality metadata in spreadsheet-based metadata acquisition systems.

**Related Work**

A variety of tools have been developed to address the limitations of spreadsheet-based metadata acquisition systems.

One of the earliest of these tools is RightField [3]. RightField is designed to improve the quality of spreadsheet-based data by guiding users to provide consistent and accurate metadata during data acquisition. It allows researchers to restrict cells to contain standardized terms from ontologies and other controlled terminologies. RightField templates can include dropdown lists and other validation rules to ensure the data entered is accurate and consistent. RightField also allows users to export collected data in various formats, including the ISA-Tab format [4], which is a standardized format for describing and sharing metadata and experimental data in life sciences research.

OntoMaton [5], which is also designed to work with the ISA-Tab format, is a similar tool that support the enforcement of the use of standard terminologies in spreadsheets. The tool allows users to search and access data from various biomedical ontologies that are present in BioPortal ontology repository [6] directly from within Google Sheets. It provides a user-friendly interface that allows users to search for terms within these ontologies and automatically retrieves related terms and associated metadata, such as definitions and synonyms. As with RightField, it is also designed to work with the ISA-Tab format and supports the creation of ISA-Tab compatible metadata templates from spreadsheets.

Mapping Master [7], adopts a different approach to the above tools. It uses spreadsheets as a starting point and provides mechanisms to maps their content to conform with ontology-based structures. It allows users to create declarative mappings between spreadsheet columns and specific terms from a given vocabulary or ontology [8]. Mapping Master uses a drag-and-drop interface to make the mapping process easy and intuitive. By mapping data to standard vocabularies, Mapping Master can quickly identify errors in spreadsheets and help to ensure that metadata are well-defined and interoperable.

Other tools address the data quality issue from a different angle, concentrating instead on providing mechanisms to repair spreadsheet-based data. One of the most popular of these types of tools is OpenRefine [9]. It provides an interface that allows users to easily clean, transform, and organize their data. With OpenRefine, users can split, merge, and reorder columns, as well as filter and manipulate rows based on various criteria. It also provides advanced features for detecting and correcting errors, removing duplicates, and transforming data into different formats. OpenRefine is widely used in data journalism, research, and business analytics, and it can handle data in various formats, including CSV, TSV, Excel, JSON, XML, and Google Sheets. Its extensibility through plugins and scripting allows for even more advanced data manipulation and integration with other tools.

While powerful, some expertise is required when using these tools and many have steep learning curves. In many situations, approaches that can be used by non-specialists when ensuring quality in spreadsheet-based content are desirable. Ideally, these interfaces should be able to quickly identify errors and to suggest repairs. An additional goal includes support for comma- and tab-separated files (CSVs and TSVs, respectively), since these remain common formats in some biomedical domains. The ultimate goal is to reduce need for human curation.

In this paper, we describe technologies we have developed how we tackle this problem. We show how these technologies were deployed in a biomedical consortium called HuBMAP to support the creation and submission of spreadsheet-based metadata templates.

**Implementation**

The technologies described in this paper were driven by the needs HuBMAP (Human BioMolecular Atlas Program)[10], a research initiative that aims to create a framework for mapping the human body at single-cell resolution. These needs drove the development of spreadsheet-focused extensions to an existing metadata management system that we developed called CEDAR.

*HuBMAP*

HuBMAP's goal is to accelerate the development of tools and techniques for constructing high-resolution spatial tissue maps of the human body and to establish an open data platform for sharing this knowledge. A key goal of HuBMAP is to make data FAIR by requiring that all submitted data metadata to adhere to community standards. The HuBMAP consortium has created a robust ingestion pipeline to help ensure this standards adherence. Using this pipeline, users can submit spreadsheet-based metadata along with associated experimental data.

Several dozen metadata specifications were developed by the consortium, with the majority of these specifications targeting a particular type of experimental technique. These specifications included information about the donors and samples from which experimental data were derived and an array of detailed specifications describing metadata for a variety of experimental assays. Each specification described metadata at the field level and indicated such things as the datatype of a field (e.g., numeric, Boolean, string) and constraints on those values (e.g., allowed value ranges for numeric fields). Sets of allowed values were also developed for many fields. These specifications were declaratively encoded using a custom YAML-based format. In a final step, spreadsheets were generated from these specifications, with each field in the specification becoming a column header in the generated spreadsheet. Both TSV and Excel formats were supported.

HuBMAP data submitters would select the appropriate spreadsheet for the type of experimental data they were submitting and then populate it with their metadata. They would then submit that metadata along with their data to HuBMAP. The ingestion pipeline analyzed the submitted metadata to ensure compliance with the associated YAML-based metadata specification. If errors were identified in the submission the ingest process was halted and error information was presented to the user. Users could then fix those errors and resubmit. While this process eventually converged to a correct submission, a significant number of iterations were often needed. In many cases, manual intervention is needed to help users supply the correct metadata.

There was thus a pressing need for a solution allowed users to quickly identify and repair errors in their metadata to ensure strong enforcement of metadata quality, while continuing to support spreadsheet-based metadata submission. Driven by these needs, we developed a series of extensions to a metadata management system called CEDAR. Our solution provides an end-to-end approach for creating, managing, acquiring, validating, and submitting high-quality spreadsheet-based metadata.

*The CEDAR System*

CEDAR The Center for Expanded Data Annotation and Retrieval (CEDAR)[1] was established in 2014 to create a computational ecosystem for the development, evaluation, use, and refinement of biomedical metadata. CEDAR supports a workflow for metadata management that is organized around the three main stages of the metadata submission process—namely, metadata specification, metadata acquisition, and metadata submission. A driving goal of CEDAR is to provide highly configurable tools that support the creation of metadata-submission pipelines to meet a wide range of deployment scenarios. It is a modular system that provides components that can be integrated into existing workflows to address specific tasks in a metadata submission pipeline or that can be assembled to provide an end-to-end pipeline. The system—referred to as the CEDAR Workbench [11]—is built around the notion of creating *templates* that define the structure and semantics of metadata specifications. These templates support a metadata-submission workflow that acquires conforming metadata and uploads the resulting metadata to repositories [12].

CEDAR's overall metadata workflow comprises the following three steps: (1) Template authors use a CEDAR tool called Template Designer to create templates describing metadata, typically following discipline-specific standards that describe the type of experimental data to be annotated. Authors can define their templates using controlled terms, ontologies, and standard datasets supplied by the BioPortal ontology portal [6]. (2) When a scientist or other metadata provider chooses to populate a template, a CEDAR tool called the Metadata Editor automatically generates a form-based interface from the template; the scientist then uses the Metadata Editor interface to enter the descriptive metadata. When users populate these forms, the semantic annotations specified by the associated template are used to present ontology-controlled suggestions to users and ensure that collected metadata conform to the published specification. (3) Once the metadata have been entered, scientists can use the CEDAR Submission Service to upload metadata and associated experimental data to a target repository

*Extending CEDAR to Support Spreadsheets*

CEDAR's capabilities were employed to provide a robust, end-to-end spreadsheet-based metadata management solution for HuBMAP.

The first step was to replace the YAML-based metadata specifications with CEDAR's template-based approach. Instead of developing YAML-based specifications, HuBMAP curators used CEDAR's Template Designer to collaboratively develop new metadata specifications. Efforts were made to expand the use of controlled terminologies in these templates to increase the quality of the resulting metadata. When possible, existing terms were chosen from BioPortal to provide values for controlled fields in metadata specifications.

While CEDAR's Metadata Editor could in principle provide metadata acquisition interfaces for each new HuBMAP metadata template, most HuBMAP consortium members use metadata creation processes that are built around spreadsheets. In some cases, these metadata are produced in spreadsheets form from computational pipelines. Supporting spreadsheet-based submission was thus crucial to ensure adoption within this community. To address this issue, we developed a mechanism to generate Excel- and TSV-based metadata spreadsheets from CEDAR templates. As with the initial HuBMAP approach, these spreadsheets provide a blank template that end users can use to submit their metadata. These spreadsheets can then be populated with correctly formatted metadata by end users. The spreadsheet generation process embeds metadata in each spreadsheet that indicate the source CEDAR templates that was used to generate it.

An interactive Web-based application was then developed to ensure that the acquired metadata adhere to the source template specification [13]. Users use this application to interactively upload and then validate their spreadsheet-based metadata (Figure 1).

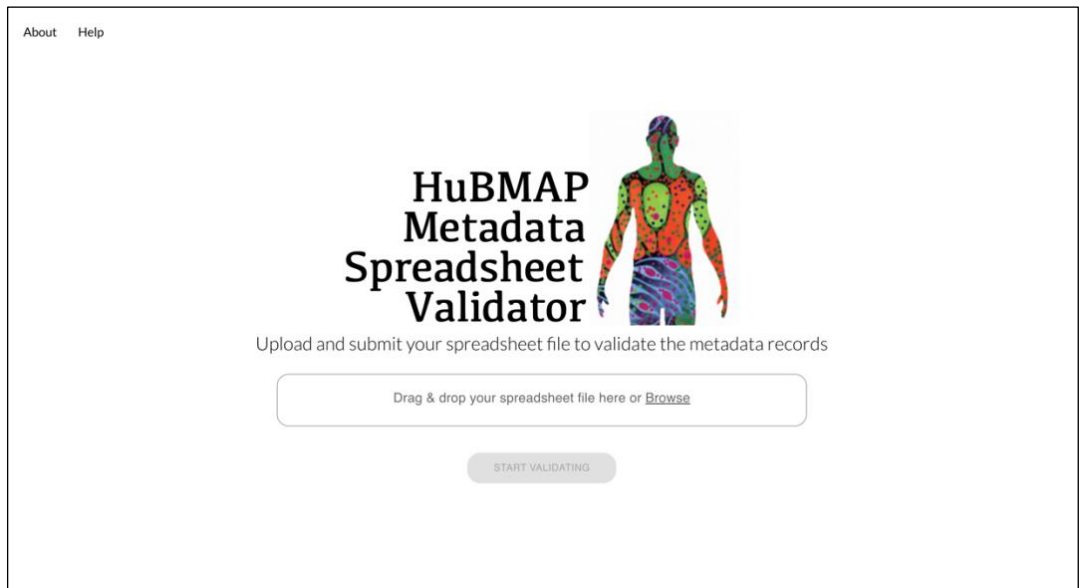

**Figure 1:** Screen shot of opening page of spreadsheet validator tool. Users can either drag-and-drop the file from their local computer to the input field or use the Browse option to select a metadata spreadsheet from their file system. Once uploaded, a *Start Validating* option is enabled

When users upload a metadata spreadsheet, they are presented with a validation dashboard. This page displays a summary of errors that were detected by the tool's algorithms.

This application adopts an array of strategies to ensure that the acquired metadata adhere to the source template specifications. It includes several wizard-style interfaces that focus on reporting validation errors in supplied spreadsheets and then on helping users to quickly repair those errors. Two primary types of validation errors are addressed: *completeness* errors and *adherence* errors (see Figure 2).

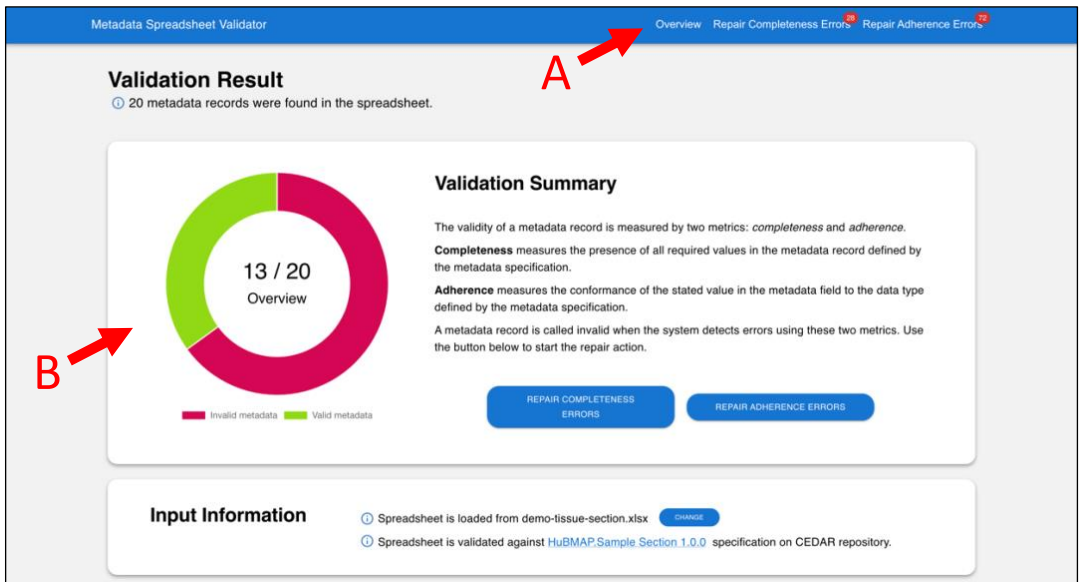

**Figure 2:** Screenshot showing the validator dashboard. The dashboard displays a summary of the errors detected in the uploaded metadata spreadsheet. These errors are divided into two types: completeness errors and adherence errors. Users can navigate to two separate wizards to repair these errors (A). A bar graph (B) indicates the total number of errors present.

Completeness validation primarily concentrates on identifying missing required values, whereas adherence validation aims to identify values that do not conform to specified value sets. The applications groups similar type of validation errors together and presents a high-level visual summary of these groups so that users can quickly spot patterns in their errors (Figure 3).

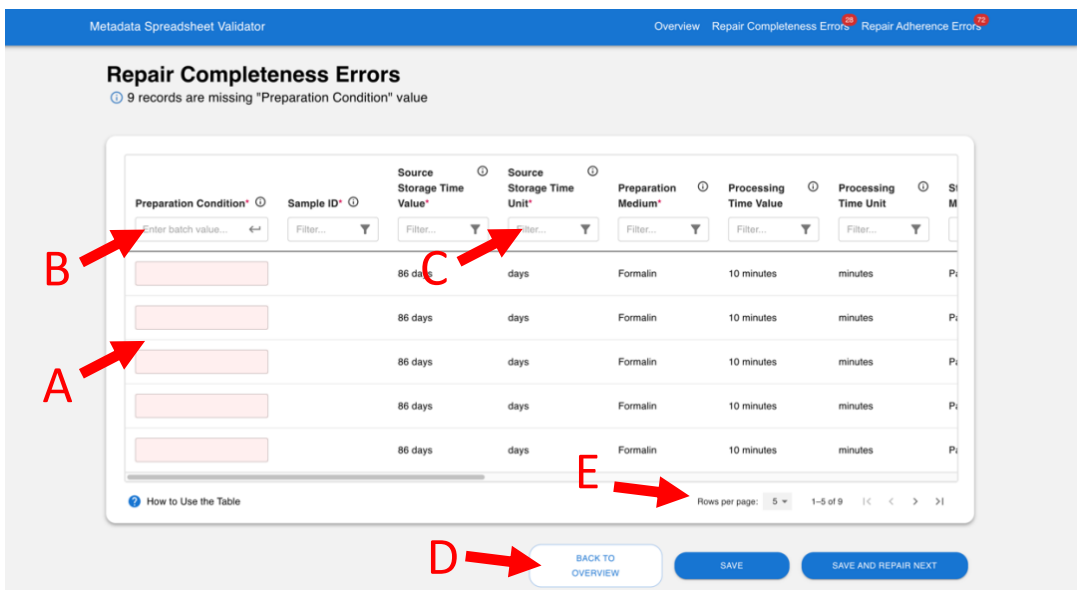

**Figure 3:** Screen shot showing the validator's wizard for repairing completeness errors. Errors are paginated and presented in a tabular form. On the left, a column (A) is presented that can be used to enter corrected values. If the value comes from a list of permissible values, then a drop-down menu will replace the standard input text field. Batch repair of rows is also possible (B). In the case of batch repair, the corrected value is applied to all the displayed rows. Field filtering (C) is also possible. At any point, users may return to the summary page (D) to review the status of the repair process. Users can also navigate to additional pages of errors (E).

A second set of interfaces then allows users to repair these errors: the first supports repair of completeness errors (Figure 4), and the second supports repair of adherence errors (Figure 4). Again, the focus is on allowing similar types of validation errors to be grouped together so that they can be repaired in batches. For each repair, an analysis is performed to identify the most likely correct values and the resulting suggestions are presented to the user, allowing users to quickly fix erroneous metadata. The suggested repairs to be applied in batches if desired. Once the repair phase is completed, users can download the repaired spreadsheet.

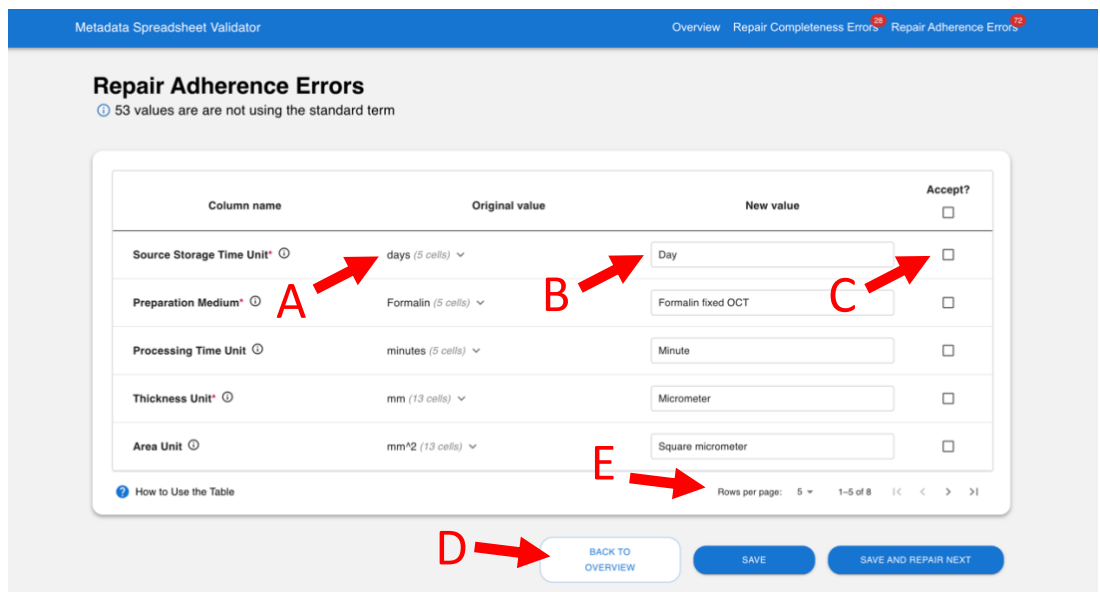

**Figure 4:** Screen shot showing the validator's wizard for repairing adherence errors. Again, the errors are displayed in tabular form. For each erroneous row the original value is presented (A). Users can directly supply new correct values or can select from suggestions presented by the wizard (B). Users can immediately accept the presented suggestions for a row (C). At any point, a summary page can be presented (D). Users can also navigate to additional pages of errors (E).

A key focus is on assisting users in quickly repairing errors by presenting suggestions for possible repairs. Simple suggestions involve inferring likely values for supplied values. For example, if the allowed values for a field are "Year", "Month", "Day" and a user supplies the value "days", the system can suggest "Day" as the corrected value using a simple string distance metric. More advanced approaches exploit the ability of CEDAR to specify controlled term value for a field. Instead of simply looking at string-based values, the validator can use the template specification for that field to query BioPortal to get all possible values for a field together with synonyms of those values. Using this expanded list, the validator can effectively exploit the semantic information in field specifications.  Other repairs supported by the validator include simple typographical repairs, such as, for example, removing quotes around numbers if they were erroneously supplied for numeric fields.

Finally, a parallel REST service was also developed to provide the necessary validation and repair functionality for the Web-based application [14]. The APIs provided by this REST service can be used by third parties to generate validation reports and repair suggestions for spreadsheets.

**Deployment**

This new metadata template-driven approach was adopted by the HuBMAP community. As mentioned, the existing YAML-based metadata specifications were replaced by metadata specifications defined using CEDAR templates. Initial donor and sample metadata specifications were replaced with CEDAR-based specifications and an array of new assay specifications were developed. When a specification is finalized we generated a human readable rendering of the specification and then publish that on HuBMAP's Web site. We also generate Excel- and TSV-based representations of each specification. These are also published on the HuBMAP web site and are available for download by metadata submitters. These submitters can populate these spreadsheets and then validate them using the validation tool. Then can then submit their metadata and associated raw data files using HuBMAP's submission processes. These processes have a parallel validation approach for validating the data supplied by submitters.

The existing HuBMAP submission processes were modified to handle handle these Excel- and TSV-based specifications. Since there is no guarantee that submitters have pre-validated their metadata using the validation tool, the HuBMAP submission system re-validates the metadata. This validation is performed by calling the validation tool's REST APIs, thus ensuring that the validation mechanisms are identical. If errors are identified in a submission, it is rejected and the user is encouraged to validate their metadata using the validation tool.

This updated submission system went live in August, 2023 and is currently in a active use. New metadata specifications are being developed and added to the submission system.

Further work could also include enhancing the validation by linking several submissions from users or groups (e.g., linking sample spreadsheets with previously validated donor spreadsheets, or assay spreadsheets with previously validated sample spreadsheets). Such an approach could also use CEDAR's user management and collaboration capabilities to link submissions together. Paragraph on future intelligent validation methods

Crucially, these interfaces should be driven by the metadata specifications used by the relevant scientific communities since conformance to those standards has increasingly become central to metadata quality assurance.

**Conclusions**

This paper has demonstrated the successful use of CEDAR technologies to provide metadata specification, acquisition, validation, and repair capabilities in the HuBMAP project. Through the collaborative development environment offered by CEDAR, high-quality metadata templates were created for describing HuBMAP donor, sample, and assay metadata. These specifications were then automatically converted into spreadsheets, which users could use to sub it their metadata to HuBMAP. The interactive Web-based service provided by CEDAR enabled users to validate and repair metadata supplied using these spreadsheets, ensuring that the acquired metadata met the quality specifications provided in the source templates. These capabilities offer a robust, end-to-end metadata management solution that addresses all stages of the HuBMAP metadata lifecycle, enabling users to quickly fix erroneous metadata and submit a repaired spreadsheet along with associated data. Overall, the use of CEDAR technologies has greatly enhanced the metadata management capabilities in the HuBMAP project.

The work provides an exemplar of how we can enhance the simple, familiar data-entry instruments that scientists want to use with features that help them to generate the kind of rich metadata that the world needs for data FAIRness. These general capabilities offer a robust, end-to-end metadata management solution that addresses all stages of the metadata lifecycle, enabling users to quickly fix non-adherent metadata and to submit a repaired spreadsheet along with associated data. Our findings demonstrate the effectiveness of our approach in ensuring compliance and quality of metadata, which can improve data sharing and reuse in scientific research.

**Citations**

[1] Wilkinson, M., Dumontier, M., Aalbersberg, I. et al. The FAIR Guiding Principles for scientific data management and stewardship. Scientific Data 3, 160018, 2016

[2] Musen MA, Bean CA, Cheung K-H, Dumontier M, Durante KA, Gevaert O, Gonzalez-Beltran A, Khatri P, Kleinstein SH, O'Connor MJ et al. The Center for Expanded Data Annotation and Retrieval. Journal of the American Medical Informatics Association, 22 (6), 2015

[3] Jupp, S., Horridge, M., Iannone, L. et al. Populous: a tool for building OWL ontologies from templates. BMC Bioinformatics 13 (Suppl 1), S5, 2012

[4] Rocca-Serra P, Brandizi M, Maguire E, Sklyar N, Taylor C, Begley K, Field D, Harris S, Hide W, Hofmann O, Neumann S, Sterk P, Tong W, Sansone SA: ISA software suite: supporting standards-compliant experimental annotation and enabling curation at the community level. Bioinformatics, 26(18), 2010

[5] Eamonn Maguire, Alejandra González-Beltrán, Patricia L. Whetzel, Susanna-Assunta Sansone, Philippe Rocca-Serra, OntoMaton: a Bioportal powered ontology widget for Google Spreadsheets, Bioinformatics, 29(4), 2013

[6] Noy, N.F., Shah, N.H., Whetzel, P.L., et al.: BioPortal: ontologies and integrated data resources at the click of a mouse. Nucleic Acids Res. 37, W170–W173, 2009.

[7] O'Connor, MJ., Halaschek-Wiener, C., Musen, MA. Mapping Master: a Flexible Approach for Mapping Spreadsheets to OWL. 9th International Semantic Web Conference (ISWC), Shanghai, China, Springer-Verlag. LNCS 6497:194-208, 2010

[8] Rocca-Serra, P., Ruttenberg, A., O'Connor, M.J., Whetzel, T., Schober, D., Greenbaum, J., Courtot, M., Sansone, S.A., Scheurmann, R., Peters, B. Overcoming the Ontology Enrichment Bottleneck with Quick Term Templates. Journal of Applied Ontology, 2012

[9] OpenRefine, https://openrefine.org/, retrieved December 11th, 2023

[10] HuBMAP Consortium. The human body at cellular resolution: the NIH Human Biomolecular Atlas Program. Nature, 574(7777):187–192, 2019

[11] Gonçalves RS, O'Connor MJ, Martínez-Romero M, Egyedi AL, Willrett D, Graybeal J, Musen MA. The CEDAR Workbench: An Ontology-Assisted Environment for Authoring Metadata that Describe Scientific Experiments. International Semantic Web Conference, 2017

[12] Egyedi AL, O'Connor MJ, Martínez-Romero M, Willrett D, Hardi J, Graybeal J, Musen MA. Using Semantic Technologies to Enhance Metadata Submissions to Public Repositories in Biomedicine. Semantic Web Applications and Tools for Health Care and Life Sciences (SWAT4LS), Antwerp, Netherlands, 2018.

[13] Spreadsheet Validator Documentation, https://metadatacenter.github.io/spreadsheet-validator-docs/, retrieved December 11th, 2023

[14] Spreadsheet Validator REST API Documentation, https://metadatacenter.github.io/spreadsheet-validator-docs/api-reference/, retrieved December 11th, 2023